# DL2Fence: Integrating <u>D</u>eep <u>L</u>earning and <u>F</u>rame <u>F</u>usion for <u>En</u>han<u>c</u>ed Detection and Localization of Refined Denial-of-Service in Large-Scale NoCs

*Haoyu Wang, Basel Halak, Jianjie Ren, Ahmad Atamli*
School of Electronics and Computer Science, University of Southampton, Southampton, UK
{haoyu.wang,basel.halak,jr1u22,a.atamli}@soton.ac.uk

## ABSTRACT
This study introduces a refined Flooding Injection Rate-adjustable Denial-of-Service (DoS) model for Network-on-Chips (NoCs) and more importantly presents DL2Fence, a novel framework utilizing Deep Learning (DL) and Frame Fusion (2F) for DoS detection and localization. Two Convolutional Neural Networks models for classification and segmentation were developed to detect and localize DoS respectively. It achieves detection and localization accuracies of 95.8% and 91.7%, and precision rates of 98.5% and 99.3% in a 16x16 mesh NoC. The framework's hardware overhead notably decreases by 76.3% when scaling from 8x8 to 16x16 NoCs, and it requires 42.4% less hardware compared to state-of-the-arts. This advancement demonstrates DL2Fence's effectiveness in balancing outstanding detection performance in large-scale NoCs with extremely low hardware overhead.

## 1 INTRODUCTION

Network-on-Chip (NoC) is a prevalent chip fabric architecture in large-scale Multi-Processor System-on-Chip (MPSoC) designs, integrating numerous IPs (Intellectual Properties) such as CPUs (Central Processing Units), GPUs (Graphic Processing Units), and memory controllers. Within the NoC, a tile, also termed a node, may function as an independent SoC which also shows NoC's flexibility and scalability. NoC-based MPSoCs primarily run applications with high-intensive on-chip computation and communication, including AI model processing and extensive simulations [1]. As a result, system performance and reliability can be easily compromised by Denial-of-Service (DoS) attacks. In [2], numerous accelerator IPs were integrated into the NoC-based MPSoC, and computation-intensive tasks were offloaded to them. During this offloading, data had to be transferred frequently, making the system highly sensitive to extraneous traffic. As the scale of a NoC increases, its performance improves, but it also requires processing more data and traffic. In such conditions, the NoC-based chip is most vulnerable to flooding attacks, a major type of DoS attacks.

DoS attacks are among the most common malicious threats in network communication, personal computers, and other large-scale chips employed in data centers. These attacks prevent users of the affected machine from accessing computational or memory resources and applications of specific IPs or modules. In the context of NoCs, a flooding attack is a type of availability attacks that aims to exhaust computing and communication resources, degrade performance, and increase unintended power consumption [3]. It often precedes hardware fuzzing attacks and confidentiality attacks. This is because the behaviour of flooding, for instance consistently sending requests to a single IP (CPU, Cache, Memory), often serves subsequent objectives. These can include, but are not limited to, destroying the victim IP or maliciously extracting information from it. Consequently, the need for flooding detection and localization in NoCs is undeniably urgent and essential for security.

DoS flooding attacks in NoCs have been modeled in several studies, such as [2, 4–10]. In these works, spurious or diverted packets were maliciously generated by malicious IPs, leading to live locks or deadlocks and finally causing a crashed system. However, only simplex consequence was given by them, more complex in the real world. We hypothesize that if the injection rate of malicious packets could be fine-tuned, a broader range of attack scenarios could be observed in more detail. In addition to its modeling, detection and localization in the NoC primarily involve monitoring network traffic, with indicators including virtual channel occupancy, buffer utilization, flit queuing time, and router latency. Traditional approaches have been explored in [4–6, 9]. With the recent advancements in machine learning, an increasing number of ML models have been utilized for DoS flooding attack detection and localization, as seen in [2, 7, 8, 10, 11]. Many of these studies have distributively integrated ML models into each router or node in a up to 8x8 NoC, achieving high-accuracy detection. However, the hardware overheads associated with these methods are significant. Inspired by this gap, we aim to introduce a novel deep learning-based multi feature frame fusion approach. This complete DoS flooding detection and localization framework delivers high-precision and accuracy through just two distinct lightweight CNN (Convolutional Neural Network) models, offering higher scalability irrespective of the NoC's scale. Our primary contributions are as follows:

- We created a Flooding DoS model with a finely adjustable injection rate, showcasing the effectiveness of attacks from malicious workloads. As a built-in function of Gem5, it can be activated and configured by input commands.
- We introduced the DL2Fence, a deep learning-based framework for DoS attack global detection and localization using a

- directional feature frame fusion technique. It enables identification of the attacking route, victims, and tracing attackers in NoC with different scales.
- We assessed the proposed framework using a refined flooding model based on and real-world PARSEC workloads and synthetic benchmarks. The hardware overhead was evaluated on real NoCs.

In this paper, Section 2 provides a background, related works, and threat modeling. Our DL2Fence scheme is introduced in Section 3. Section 4 gives us an explanation on feature and model selection. Results are evaluated in Section 5, and we conclude in Section 6.

## 2 BACKGROUND AND THREAT MODELING

### 2.1 DoS attack modeling in NoCs

As one of the most prevalent attacks, DoS can induce several adverse effects in NoC-based systems. Among the most detrimental and immediately noticeable impacts on user experience are performance degradation, diminished quality of service, and a surge in power consumption. [9] underscores significant performance degradation in MPSoCs due to DoS flooding attacks. It sheds light on different attacking scenarios, including 3 types of overlaps and loop attacks. [2] conceptualizes their flooding attack based on the number of attackers and victims by enumerating attack patterns composed by Malicious IPs (MIPs) with Victim IPs. [6] was the first to coin the term FDoS for DoS flooding attacks in NoCs, discussing attacks not instigated by Hardware Trojans (HT) but by malicious applications executed on a processing element. The study also briefly delves into two FDoS implementation scenarios: one that increases the packet injection rate and another that extends the packets' payload length. These insights underscore the diversity and intricacy of DoS attacks initiated by malicious workloads.

Nevertheless, to the best of our knowledge, a few publications examine the method of refined DoS modeling in NoCs or the nuanced impacts on a system as the injection rate of flooding packets increases from non-attacking communication modes. Especially noteworthy is their lack of trials or demonstrations of DoS overlaying on normal workloads, an approach necessary to assess the proposed security framework in a more realistic scenario.

### 2.2 DoS Detection and Localization in NoCs

There are detectable features that can be used to detect the onset of DoS attacks. However, without the aid of machine learning, certain processes and specific algorithms must be designed to make an accurate decision. Traditional approaches monitor packet-related features to detect DoS. [9] carefully tracks packet violations to detect and localize attacks in real-time. Two curves, the Packet Arrival Curve and Destination Latency Curves, were proposed as essential evidence for attack detection and were employed in a distributed attacker localization protocol. In essence, all security protocols are executed in each router during every detection period. [6] also designed DoS monitors, competitors, and updaters for each port of every router. Their scheme led to area overheads of 32%–39%. Unlike [9], their work cannot pinpoint the attacker of FDoS.

Machine learning models are increasingly being utilized for DoS detection and localization in NoCs, showcasing various advantages. [2] stands out as a prime example, having trained and integrated a perceptron model into each router. However, five additional modules had to be incorporated into each router to aid in the detection and localization of DoS. two added modules are used to localize the MIP. Their "Sniffer" achieved an area overhead of 3.92% per router. Another series of notable works on distributed ML-based attack detection comes from Ke Wang [11, 12]. In their paper [11], they trained a Generative Adversarial Network to discern between normal and HT-infected time series distributions of NoC attributes, subsequently detecting the HT. They proposed multiple mitigation strategies targeting HT to prevent the system from suffering high latency and excessive power consumption due to HT attacks. Although they achieved their desired results, the area overhead on each router was 7.2%, slightly higher than in [2]. Therefore, similar scalability concerns arise. Unlike distributed ML-based methods, [7, 8, 13, 14] leverage a single trained ML model, such as ANN (Artificial Neural Network) and XGB (XGBoost Classifier), to detect DoS attacks. This indicates that high accuracy can be attained using fewer machine learning models for global detection, as opposed to a distributed, router-based approach.

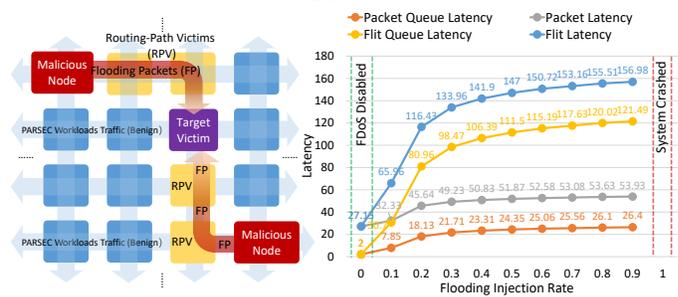

Figure 1: Flooding DoS overlaying on normal workload traffics, and increasing system latency with FIR

### 2.3 Proposed Fine-Tuning Flooding DoS

In this work, we assume that at least one malicious node consisting of at least one processing element simultaneously initiates a flooding attack on a target victim node by injecting and sending superfluous packets, such as unlimited requests or acknowledges. Meanwhile, normal communication on all nodes must not be paused or halted, but just be slowed down. In other words, the flooding attack overlays or interludes normal communication traffic, but the normal traffic efficiency will be degraded to some extent, representing an 'ideal' flooding attack implementation illustrated in the left of Figure 1. Additionally, adversaries have not compromised network protocols and routing algorithms to execute more stealthy attacks. For example, router communication credits stay balanced as all transactions, even those initiated by flooding attackers, are legitimate; malicious packets follow paths determined by system default routing algorithm. In this manner, the NoC protocol permits flooding to persist, and the system can only be overwhelmed by a deluge of malicious packets, not by unauthorized communications. As a result, system runtime features dramatically vary based solely on the FIR (Flooding Injection Rate), such as performance degradation, or system crash due to traffic congestion. It was implemented and inserted within workload applications as a malicious *'Tick'* function and features adjustable input parameters. The right figure in Figure 1 shows the latency-FIR relationship in our flooding model. Latency rises with higher FIR, heavily impacting system performance. The increment on latency from FIR 0.1 to 0.9 is 1.1 to 60 times over normal value. At FIR=1, the system crashes; lower FIR values make the model stealthier but sustain the negative impact.

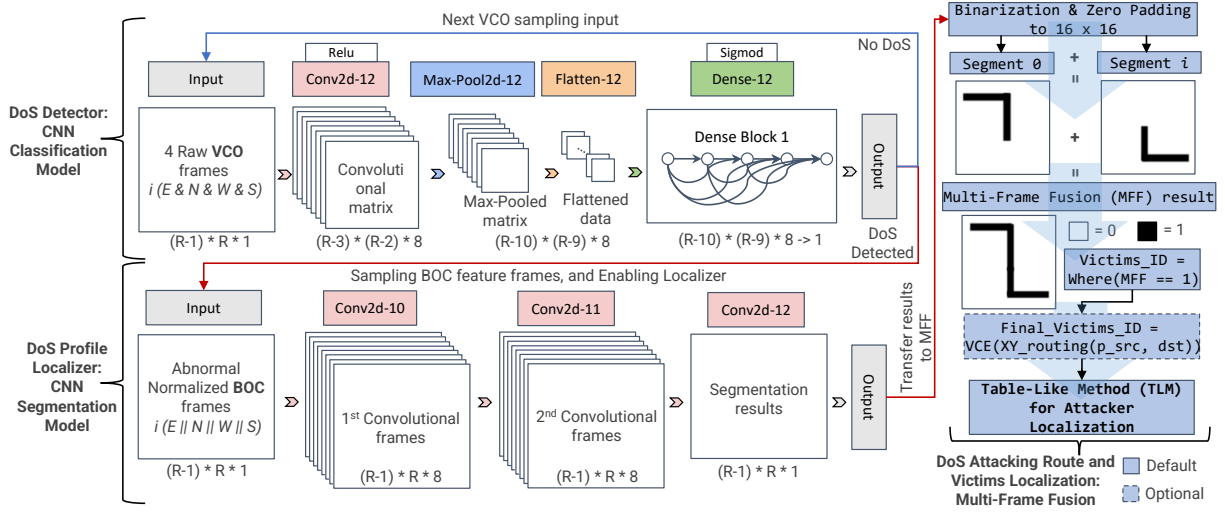

Figure 2: An overview of the proposed DL2Fence: Deep Learning based Multi-Frame Fusion Method for DoS Detection and Localization (R: Mesh Row Number)

## 3 DL2FENCE: DEEP LEARNING AND FRAME FUSION FRAMEWORK

Drawing inspiration from computer vision techniques, notably pattern recognition and semantic segmentation, we believe that the runtime global features of all routers in NoCs can be visualized as image frames. This representation allows us to approach the problem as an object detection and segmentation task within images, especially fitting for NoCs with regular topologies such as a 2D mesh or further expand to 3D mesh. As shown in figure 4, the data for Virtual Channel Occupancy (VCO) and Buffer Operation Counts (BOC) from the input ports of routers in a 16×16 NoC at a specific runtime point were extracted using the Gem5 simulator. An evident DoS attack is indicated by abnormally high local feature data in these frames. Additionally, the BOC matrix frames clearly reveal profiles of abnormal routes. These observations motivate our approach to employ CNN models for detecting and localizing anomalies in NoCs.

Figure 2 presents the comprehensive proposed framework, composed of three stages: **i. DoS Detector: CNN Classification**, **ii. DoS Profile Localizer: CNN Segmentation**, and **iii. Victims and Attackers Localization: Multi-Frame Fusion (MFF) combined with Table-Like Method (TLM)**. The operational flow of the framework can be summarized as:

(1) Periodically, the detector processes every sampled VCO feature frames from 2-4 directions of each router, provided no attack is detected before;
(2) If any data frame is identified as anomalous by the detector, all corresponding BOC feature frames are collected and sent to the segmentation model-based localizer;
(3) Merging segmented frames reconstructs the attacking routes and all victims will then be localized. The TLM-based algorithm accurately pinpoints attackers;
(4) The system then quickly proceeds to the next VCO sampling and detection/localization round, ensuring rapid identification of any attackers missed in the previous round, repeating until no abnormal frames appear.

The individual components of the framework are elaborated upon in the subsequent subsections.

### 3.1 DoS Detector: CNN Classification

A CNN classification model has been trained on VCO features and is used to detect the presence of a DoS flooding attack. The rationale for feature selection is detailed in Section 4. The model requires four frames as inputs because routers can have up to four directions of input ports. In the input layer of the DoS detector shown in figure 2, 'E & N & W & S' depict the necessary 4-directional input frames. Every router situated in the center of the NoC has four directional ports; those on the edge have three; and those in the corners have only two. Note that routers on the edges lack external NoC input ports. Thus, the feature frame in this study always forms an $R \times (R-1)$ matrix, where $R$ is the row number of a 2-D Mesh NoC. The VCO value fluctuates based on runtime traffic density and increases when more virtual channels are occupied simultaneously.

From a perspective of balancing hardware overhead and detection accuracy, a lightweight classification model has been crafted (in Figure 2). It consists of an input layer, a convolutional layer with a ReLU activator, a max-pooling layer, a flatten layer, and a dense layer that produces an output value. During its training process, the model abstracts and extracts information from the 4-directional input frames, retaining the most critical and informative values as the model's weights and biases. Consequently, when the trained model encounters similar feature frames, it can make reliable predictions.

### 3.2 DoS Profile Localizer: CNN Segmentation

Another CNN model has been trained for segmentation tasks using BOC frames, which are collected simultaneously when VCO frames exhibit abnormalities. Deep learning-based semantic segmentation techniques are extensively used in medical imaging, as exemplified by the renowned segmentation model, U-Net [15]. We leverage the high precision of CNN segmentation models to localize flooding attacks and, consequently, reconstruct the entire attacking route. The input layer of the DoS localizer, as depicted in figure 2, includes abnormal 'E || N || W || S' to indicate that not all frames are necessary as inputs, thereby reducing the computation time and

| Frame Combination / Abnormal Dir Frame | One Abnormal Frame | Two Abnormal Frames | | | | | Three Abnormal Frames | | | | Four Abnormal Frames |
|---|---|---|---|---|---|---|---|---|---|---|---|
| | | E & N/S or W & N/S | | | | E & W or N & S | E & N & W | E & W & S | E & N & S | W & N & S | E & N & W & S |
| | | (Max(N/S) − Min(N/S)) % R == 0 | | | | | | | | | |
| | | Max(E/W) - Min(E/W) < R-1 | | (2R-1, 3R-1...) > Max(E/W) - Min(E/W) > R | | | | | | | |
| Attack Path Example (Frame Fusion Results) | | | | | | | | | | | |
| Attacker Number | 1 | 1 | | ≥2 | | ≥2 | ≥2 | ≥2 | ≥2 | ≥2 | ≥2 |
| E=1 | Max(E) + 1 | Max(E) + 1 | Max(E) + 1 | Max(E) + 1 by multiple samples | Max(E) + 1 by multiple samples | Min(W) − 1, Max(E) + 1 | Max(E) + 1, Min(W) - 1 | Max(E) + 1 | Max(E) + 1, Max(N) + R | | Max(E) + 1, Min(W) − 1 |
| N=1 | Max(N) + R | | | | | | | | | Min(W) − 1, Max(N) + R, Min(S) − R | |
| W=1 | Min(W) − 1 | Min(W) − 1 | Min(W) − 1 | Min(W) − 1 by multiple samples | Min(W) − 1 by multiple samples | Min(S) − R & Max(N) + R | Min(W) - 1 | Min(W) - 1 | | | |
| S=1 | Min(S) − R | | | | | | | Min(S) - R | Min(S) − R | | |

Figure 3: Table-Like Method based Attacker Localization (E: East, N: North, W: West, S: South, R: Mesh Rows)

overhead. Feature frames, representing buffer operation counts, are gathered from every input port of the routers. These frames denote the number of buffer reads/writes and can be easily tallied.

From a similar viewpoint, we designed a fundamental segmentation model (in Figure2) comprising an input layer, two convolutional layers (1st and 2nd), and an output layer. With feedback from dice accuracy, the model can refine its parameters based on the disparities between segmentation results and ground truth frames. It is worth noting that while adding more convolutional layers might enhance dice accuracy, it would substantially inflate the model's hardware overhead in its implementation. Based on acceptable hardware overhead and required accuracy, structures can be dynamically adjusted.

### 3.3 DoS Attacking Route, Victims, and Attacker Localization

The first crucial step before attacker localization is the reconstruction of the flooding attacking route, which is composed of the routing-path victims (RPV) as shown in Figure 1. As outlined in Algorithm 1 and illustrated on the right side of Figure 2, the process of localizing victims incorporates the MFF technique. Additionally, a configurable function named Victim Complementing Enhancement (VCE) is introduced, which can refine and complete RPV localization under certain conditions. Before applying the MFF technique, it is necessary to adjust the segmentation results to a standard binary 16 × 16 matrix by binarization and padding zeros. In the MFF result frame, nodes with a pixel value 1 indicate the identified victims. Inspired by the RPV deduction technique in [14], the VCE function refines results using a pseudo-source ID adjacent to the attacker and a specified destination ID in reverse XY routing algorithm. This method can deduce all correct RPVs between two input IDs. The VCE is designed as a configurable function because it yields the best results when the initial detection phase is accurate enough.

The final stage of the framework is attacker localization. Using the XY-routing algorithm, which attackers also follow, we can determine the exact number of attack patterns. Feature frames provide directional information, leading us to propose a Table-Like Method (TLM) that enumerates all single or multi-attacker scenarios for a single victim. This approach is translated into algorithmic code and detailed in figure 3. The table's initial rows list the number of abnormal frames detected, conditions that signify different attack patterns, and combinations of abnormal directional frames. The first column delineates attacking routes, predicted attacker numbers, and indicators for abnormal directional frames (e.g., **E=1** indicates abnormal data from **east input ports**). The cells between the 4th row and 7th row hold the attacker localization results, with **Max('D')** indicating the **biggest RPV ID in direction 'D' (E, N, S, W)**. In short, attacker IDs are calculated using combination of input port direction and the XY-routing algorithm. For instance, if a single attacker floods from the east, one abnormal east VCO feature frame would be detected. With the **'East'** information, the attacker must be to the **Right** of the victim, and therefore, their ID would be the maximum ID among RPVs plus 1. We aimed to minimize localization rounds with TLM, localizing all attackers in as less cycles as possible. Extra conditions under 'Two Abnormal Frames' cover simpler patterns, while '≥ 2' scenarios require multiple rounds. In summary, multi-attacker scenarios require multiple 1-2 attacker localization rounds.

---

**Algorithm 1:** Victim (attacking route) Localization

**Data:** Segmented frames: seg_result_d (d: e, n, s, w)
**Result:** All victim node IDs including Routing-Path Victims (RPV) and Target Victims (TV)

1. **while** *Getting all seg_result_d* **do**
2.     $seg\_bin\_d = Binarization(seg\_result\_d)$
3.     $full\_frame\_d(16 \times 16) \leftarrow Zero\_Pad\_R/L/T/B(seg\_bin\_d)$
       /* Multi-Frame Fusion */
4.     **while** $i \leq Number\ of\ (malicious)seg\_result\_d$ **do**
5.        $MFF+ = full\_frame\_d[i]$
6.     **end**
7.     $victim\_x, victim\_y \leftarrow Where(MEF(pixel\_value == 1))$
8.     $victim\_id \leftarrow Get\_Node\_ID(victim\_x, victim\_y)$
       /* VCE by xy-routing deducing */
9.     **if** *Enable Victim Completing Enhancement* **then**
10.        $pseudo\_src \leftarrow Get\_SRC(vic\_id)$
11.        $TV \leftarrow dst \leftarrow Get\_DST(vic\_id)$
12.        $RPV \leftarrow XY\_Routing(pseudo\_src, dst)$
13.     **end**
14. **end**

---

## 4 FEATURES AND MODEL SELECTION

[7] utilized average link utilization for model training, showing its sensitivity to DoS attacks. However, this 'external' feature fails to capture router-specific data, thereby lacking directional information. In contrast, we opted for VCO of routers' input ports to offer directional insights and virtual channel status. VCO is an instantaneous value ranging from 0 to 1, derived from the ratio of occupied VCs to total VCs in an input port. For a more comprehensive view,

we included an accumulated feature, BOC, which accounts for the number of buffer writes/reads.

The suitability of VCO and BOC for detection or localization is determined through hardware design trade-off and experimental results. The framework with VCO excels in detection, but fares poorly in localization (traffic-intensive benchmarks). VCO's floating-point nature allows for excellent detection performance even without normalization. However, its localization efficacy is limited due to incomplete attacking route observation, as exemplified in figure 4. BOC, on the other hand, performs admirably in both detection and localization, with higher accuracy. However, BOC requires normalization due to its integer type. Thus, to minimize computational overhead, VCO is chosen for detection and BOC for localization. This approach allows for direct utilization of VCO without normalization and necessitates BOC normalization only when localization is required, satisfying accuracy requirements for both stages. For less traffic-intensive workloads, such as PARSEC, employing VCO for dual tasks is feasible, as elaborated in Section 5.2.

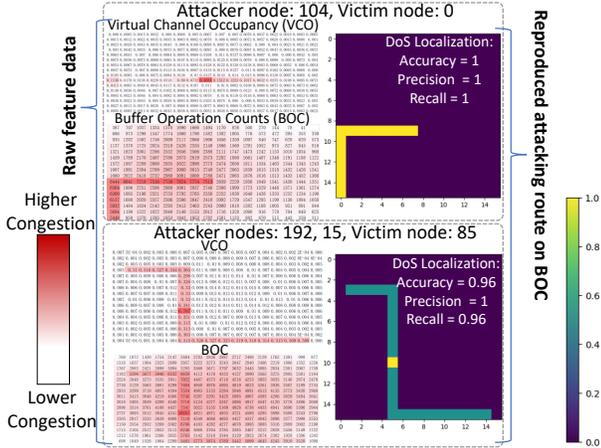

Figure 4: Localization examples (on STP benchmark)

Initially, we chose a simple-structured CNN model for these tasks, given its widespread deployment as customized accelerators and proven effectiveness. Although BNN (Binarized Neural Network) models have shown good performance and low overhead [16], our attempts to train a BNN model on BOC and VCO features were unstable compared to the CNN model. We experimented CNN models with various architectures, including altering the number of filters and adjusting layers. While these changes resulted in marginal accuracy gains, the hardware and power overhead they introduced extremely outweighed the benefits. As a result, we opted for two minimal CNN models with 8 convolution kernels in each layer, as depicted in Figure 2.

## 5 EXPERIMENTS AND RESULTS

The threat model and framework were firstly implemented in Garnet [17] within the Gem5 [18] simulator. A 16 × 16 Mesh-XY NoC, which is larger than scales attempted in all related works, was constructed with one VNET and default virtual channels. Six synthetic traffic pattern (STP) benchmarks and three PARSEC workloads were ran to evaluate proposed approach and the flooding model. However, due to Gem5 simulator limitations, PARSEC can only be run on an NoC with a maximum size of 8x8. The system clock was set at 2 GHz, with features sampled every 1000 cycles for STP,

100000 cycles for PARSEC. Higher system frequencies could allow shorter monitoring cycles, depending on the SoC's sensitivity of security requirements. We designed a global performance monitor to collect the dataset, simulating 18 attack scenarios under 0.8 FIR across 6 + 3 benchmarks, yielding 162 datasets. Totally 12960 feature frames were extracted from four directions of input ports. With 256 pixels per frame, the dataset contains a total of 3.3 million pixel values. The TensorFlow 2.0 framework was imported to do off-line training of CNN models. Lightweight CNN detection and localization accelerators and ProNoC[19]-generated NoCs were synthesized to evaluate the hardware overhead of the proposed method. We initially tested our framework for both DoS detection and localization on each of VCO and BOC (see tables 1, 2) to find the best feature combination. The results of the chosen VCO+BOC feature set are in table 3.

### 5.1 Detection Performance

The detection performance of CNNs using the VCO feature is presented in the left numbers of each column in table 1, showing an average accuracy of 0.98 and precision of 0.99 on STP but 0.93 and 0.96 on PARSEC, highlighted in the last column. The BOC feature's detection performance, detailed in table 2, achieves a robust average accuracy of at least 0.99 and a precision of 1. VCO, despite its slightly lower performance compared to BOC due to less pronounced variations, is chosen for detection tasks as it does not need normalization for model inference and its instantaneous value uses less memory than BOC's accumulating value.

### 5.2 Localization Performance

Localization performance of STP running on VCO feature underperforms compared to BOC, with an average gap of 0.45 in accuracy, which can be found in the right of each column in table 1. BOC features are therefore more effective for localization in STP benchmark, as evidenced by the framework's high metrics: accuracy 0.973 and 0.917 in table 2 and table 3. Specific localization examples of BOC data are provided in figure 4, in which BOC data showcases a more clear attacking routes. Localization tasks perform well on the PARSEC benchmark, largely attributed to the lower traffic density during the Region-of-Interest (ROI) period, where more computations occurred instead. Consequently, the FDoS was readily detected and localized within the ROI. For PARSEC workloads, exploring the use of VCO for simultaneous detection and localization presents a promising scheme which will be good at less-traffic-intensive tasks.

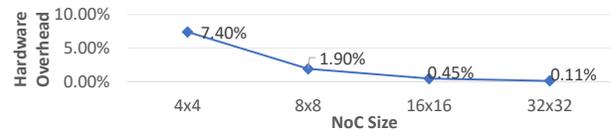

Figure 5: Hardware overhead decreasing with larger NoCs

### 5.3 Hardware Overhead and Related Works

Two CNN accelerators were developed with minimized logic usage, incorporating three convolutional kernels in a pipeline architecture. Their hardware overhead was assessed on a complete NoC [19], comprising only routers, network interfaces and links, excluding SoC tiles to eliminate inaccuracies. The models showed overheads of 1.9% and 0.45% in 16x16 and 8x8 NoCs, respectively, indicating a 76.3% decrease in overhead as NoC size increases. The trend of decreasing hardware overhead is visible in Figure 5, highlighting our framework's adaptability, efficiency, and lightweight nature for various SoC needs. At the 8x8 scale, our design achieves 42.4% less hardware overhead compared [2], despite a slightly larger model than the perceptron model. A performance comparison with other works is provided in Table 4, including selected model specifications, hardware overhead, and the detection matrix, all focusing on DoS attack detection, particularly flooding attacks [2, 8]. Our work is unique in presenting results for up to 16x16 NoCs, demonstrating greater scalability and reduced hardware overhead. In contrast, other distributed schemes have a constant hardware overhead regardless of NoC scale. While our accuracy is average, our precision in detection and localization is superior to others. DL2Fence's focus on traffic-heavy STP benchmarks for comparison provides stricter NoC communication performance assessments, as conservative comparison. However, it excels in PARSEC benchmarks, as shown in the last four columns of Tables 1 and 2, where less data exchange but more computations are needed, making the flooding traffic more prominent. Thus, DL2Fence likely surpasses state-of-the-arts in realistic workloads.

Table 1: DoS Detection and Localization Results for Virtual Channel Occupancy Feature (without normalization)

| Benchmarks: | Detection results (left) \| Localization results (right) on VCO feature matrix (1 attacker + 2 attackers together) | | | | | | | | | | |
|---|---|---|---|---|---|---|---|---|---|---|---|
| | Synthetic Traffic Pattern (STP) | | | | | | | PARSEC | | | |
| Metric | Uniform Random | Tornado | Shuffle | Neighbor | Bit Rotation | Bit Complement | Average | Blackscholes | Bodytrack | X264 | Average |
| Accuracy | 1\|0.74 | 0.96\|0.43 | 0.93\|0.31 | 0.97\|0.61 | 1\|0.58 | 1\|0.52 | **0.98**\|0.53 | 0.95\|0.98 | 0.92\|0.99 | 0.94\|0.99 | **0.93**\|**0.98** |
| Precision | 1\|0.89 | 1\|0.63 | 0.97\|0.42 | 0.98\|0.84 | 1\|0.68 | 1\|0.65 | **0.99**\|0.69 | 1\|0.99 | 1\|0.99 | 0.89\|1 | **0.96**\|**0.99** |
| Recall | 1\|0.8 | 0.92\|0.52 | 0.92\|0.56 | 0.97\|0.65 | 1\|0.79 | 1\|0.68 | **0.96**\|0.67 | 0.92\|0.99 | 0.87\|0.99 | 0.98\|1 | **0.92**\|**0.99** |
| F1 Score | 1\|0.84 | 0.95\|0.54 | 0.93\|0.44 | 0.97\|0.71 | 1\|0.72 | 1\|0.65 | 0.97\|0.65 | 0.96\|0.99 | 0.93\|0.99 | 0.93\|0.99 | **0.94**\|**0.99** |

Table 2: DoS Detection and Localization Results For Buffer Operation Counts Feature (with normalization)

| Benchmarks: | Detection results (left) \| Localization results (right) on BOC feature matrix (1 attacker + 2 attackers together) | | | | | | | | | | |
|---|---|---|---|---|---|---|---|---|---|---|---|
| | Synthetic Traffic Pattern (STP) | | | | | | | PARSEC | | | |
| Metric | Uniform Random | Tornado | Shuffle | Neighbor | Bit Rotation | Bit Complement | Average | Blackscholes | Bodytrack | X264 | Average |
| Accuracy | 1\|0.97 | 1\|0.98 | 1\|0.97 | 1\|0.98 | 0.99\|0.97 | 0.99\|0.97 | **0.997**\|**0.973** | 0.93\|0.94 | 0.99\|0.99 | 0.92\|1 | **0.94**\|**0.97** |
| Precision | 1\|1 | 1\|1 | 1\|1 | 1\|1 | 1\|1 | 1\|1 | **1.000**\|**1.000** | 1\|0.99 | 1\|1 | 0.84\|1 | **0.94**\|**0.99** |
| Recall | 1\|0.97 | 1\|0.98 | 1\|0.97 | 1\|0.98 | 0.97\|0.97 | 0.97\|0.97 | **0.990**\|**0.973** | 0.85\|0.99 | 0.96\|1 | 1\|1 | **0.93**\|**0.99** |
| F1 Score | 1\|0.99 | 1\|0.99 | 1\|0.98 | 1\|0.99 | 0.98\|0.98 | 0.98\|0.98 | 0.993\|0.985 | 0.92\|0.99 | 0.98\|1 | 0.91\|1 | **0.93**\|**0.99** |

Table 3: DoS Detection and Localization Results For Virtual Channel Occupancy + Buffer Operation Counts Features

| Benchmarks: | Detection results on VCO (left) \| Localization results on BOC (right) feature frames | | | | | | | | | | |
|---|---|---|---|---|---|---|---|---|---|---|---|
| | Synthetic Traffic Pattern (STP) | | | | | | | PARSEC | | | |
| Metric | Uniform Random | Tornado | Shuffle | Neighbor | Bit Rotation | Bit Complement | Average | Blackscholes | Bodytrack | X264 | Average |
| Accuracy | 0.99\|0.96 | 0.93\|0.84 | 0.93\|0.86 | 0.9\|0.9 | 1\|0.97 | 1\|0.97 | **0.958**\|**0.917** | 0.95\|0.89 | 0.92\|0.91 | 0.94\|0.94 | **0.933**\|**0.913** |
| Precision | 1\|1 | 1\|1 | 0.97\|0.96 | 0.94\|1 | 1\|1 | 1\|1 | **0.985**\|**0.993** | 1\|0.99 | 1\|1 | 0.89\|0.89 | **0.963**\|**0.960** |
| Recall | 0.98\|0.96 | 0.86\|0.84 | 0.92\|0.89 | 0.85\|0.9 | 0.99\|0.97 | 0.99\|0.97 | **0.932**\|**0.922** | 0.92\|0.91 | 0.87\|0.87 | 0.98\|0.98 | **0.923**\|**0.920** |
| F1 Score | 0.99\|0.98 | 0.9\|0.89 | 0.92\|0.9 | 0.88\|0.93 | 1\|0.98 | 1\|0.99 | 0.948\|0.945 | 0.96\|0.94 | 0.93\|0.93 | 0.93\|0.93 | **0.940**\|**0.933** |

Table 4: Comparison to related works

| Works | [2] | [13] | [8] | [Our Work] |
|---|---|---|---|---|
| ML Model(s) | Perceptron | SVM | XGB | CNN Classifier + Segmentor |
| FDoS | ✓ | Non-FDoS | ✓ | ✓ |
| Scalability | ✓ | ✓ | ✓ | ✓ |
| Hardware Overhead | 3.3% | 9% | N/A | 0.45%/16x16NoC 1.9%/8x8NoC |
| DoS Detection & Localization Performance | | | | |
| NoC Scale | 8x8 | 4x4 | 4x4 | 16x16 ★ |
| D: Accuracy | 97.6% | 95.5% | ~96% | **95.8%** |
| D: Precision | N/A | ~94.5% | 94.8% | **98.5%** |
| L: Accuracy | 96.7% | N/A | N/A | **91.7%** |
| L: Precision | N/A | N/A | N/A | **99.3%** |

**D**: Detection. **L**: Localization

## 6 CONCLUSION

This paper introduces DL2Fence, a deep learning-based framework for Flooding DoS (FDoS) detection and localization, featuring a proposed fine-tuning FDoS model. Our framework utilizes two CNN models for effective feature frame classification and segmentation, and to respectively compose DoS detector and DoS profile localizer. It also integrates techniques such as Multi-Frame Fusion, Victim Completing Enhancement, and a Table-Like Method for precise localization. DL2Fence achieves a detection accuracy of 95.8% and a localization accuracy of 91.7% on a 16*16 mesh level, with a precision of approximately 99% in both aspects, comparable to state-of-the-art solutions. Significantly, it reduces hardware overhead by 76.3% when transitioning from 8x8 to 16x16 NoCs. The framework is scalable for larger NoCs, ensuring optimal performance with minimal overhead. For NoCs larger than 32x32 nodes, a MobileNet CNN model is employed to further improve FDoS profile localization, incurring an estimated hardware overhead of no more than 2.5%.